\documentclass[sigconf, screen]{acmart}
\usepackage{url}
\usepackage{graphicx}
\usepackage{booktabs}
\usepackage{makecell}
\usepackage{multirow}
\usepackage{textcomp}
\usepackage{flafter}
\usepackage{stfloats}
\usepackage{csquotes}
\usepackage{soul}
\usepackage{float} 
\usepackage{siunitx}
\usepackage{array}
\newcolumntype{L}[1]{>{\raggedright\arraybackslash}p{#1}}
\AtBeginDocument{%
  }

\setcopyright{acmlicensed}
\copyrightyear{2018}
\acmYear{2018}
\acmDOI{XXXXXXX.XXXXXXX}
\acmConference[Conference acronym 'XX]{Make sure to enter the correct
  conference title from your rights confirmation email}{June 03--05,
  2018}{Woodstock, NY}
\acmISBN{978-1-4503-XXXX-X/2018/06}

\newcommand{\XDCMT}[1]{{\leavevmode\color{black}{#1}}} 
\newcommand{\BLCMT}[1]{{\leavevmode\color{black}{#1}}} 

\begin{document}

\title{Privacy Perceptions in Sensor-Powered Smart Vehicle Cabins}
\author{BoRui Li}
\email{borui_li@sfu.ca}
\orcid{0009-0005-2730-4471}
\affiliation{%
  \institution{Simon Fraser University}
  \city{Burnaby}
  \state{British Columbia}
  \country{Canada}
}
\author{Bofan Yu}
\email{bofany@sfu.ca}
\affiliation{%
  \institution{Simon Fraser University}
  \department{School of Computing Science}
  \city{Burnaby}
  \state{British Columbia}
  \country{Canada}
}

\author{Xing-Dong Yang}
\email{xingdong_yang@sfu.ca}
\affiliation{%
  \institution{Simon Fraser University}
  \city{Burnaby}
  \state{British Columbia}
  \country{Canada}
}

\renewcommand{\shortauthors}{Trovato et al.}



\begin{abstract}
As car cabins evolve with the integration of diverse sensors, traditional car cabins are transforming into smart environments. This shift raises important questions about how privacy is understood and managed in such spaces. In this work, we investigate privacy perceptions from the perspectives of both vehicle owners (i.e., people who purchase and own cars) and non-owners (i.e., people who temporarily use cars, such as family members, friends, or renters). Through semi-structured interviews with eighteen participants, we identified key factors that influence these groups’ views on privacy. Our findings reveal factors that commonly influence privacy preferences for both owners and non-owners, as well as factors that have a stronger impact on one group over the other. Drawing on these insights, we discuss design implications for future designs to better support and balance the diverse privacy needs of multiple stakeholders in smart car cabins.
\end{abstract}

\ccsdesc[500]{Human-centered computing~Empirical studies in HCI}
\keywords{Privacy, Interview, Smart Vehicle Cabins}
\maketitle

\section{Introduction}
Modern automobiles now incorporate a wide range of sensors. While some, such as wheel speed sensors, primarily support vehicle operation, many in-cabin sensors (hereafter referred to as \textit{smart cabin sensors}) are designed to capture driver and passenger activities with the aim of enhancing both safety and user experience \XDCMT{\cite{Mishra, Lu}}. Beyond well-known visual and auditory signals collected through cameras and microphones, smart cabin systems can also sense a variety of physiological indicators, such as heart rate and breathing rate \XDCMT{\cite{Gong, Hou}}. These diverse data streams provide access to rich behavioral and biometric information, often without users' awareness \XDCMT{\cite{Leonhardt, Kim3}}.

Much of the existing research on automotive privacy has focused on technical approaches, such as preventing the sharing or misuse of sensitive data \XDCMT{\cite{Duri,Babaghayou,Pape}}. In contrast, less is understood about how privacy preferences vary across everyday driving contexts. These preferences may be shaped by several factors. For example, social dynamics can influence privacy expectations, which may vary when driving alone versus with passengers. Similarly, car ownership can affect users’ sense of control and privacy needs, leading to different expectations for vehicle owners compared to non-owners. Understanding how privacy preferences shift across these dynamic contexts is important for designing smart cabin sensing systems that can adapt to and respect users’ privacy needs.

Previous work has examined car users’ privacy preferences in specific scenarios, often focusing on drivers or vehicle owners traveling alone, framing the cabin as a private space \XDCMT{\cite{Dowthwaite}}. Our work builds on this foundation by exploring more common and diverse contexts. In particular, we examine the factors shaping privacy preferences across two key stakeholder groups: vehicle owners (i.e., individuals who purchase and own the car) and non-owners (i.e., individuals who use the car temporarily, such as family members, friends, or renters). We use this distinction as an initial comparative lens to examine how privacy preferences shift with users’ relationship to the vehicle. We further broaden the scope of investigation by considering a variety of situations, including instances where owners travel alone or with others, both as drivers and passengers, as well as cases where non-owners occupy the vehicle under similar conditions. Guided by this perspective, our study addresses two research questions: (1) What factors influence car users’ privacy preferences? (2) How do these factors differ between vehicle owners and non-owners? 

To explore these questions, we conducted semi-structured interviews with 18 regular car users. Our findings reveal that, beyond previously reported factors such as data type and cost–benefit trade-offs \XDCMT{\cite{Dowthwaite, Walter}}, additional considerations shape privacy preferences for both owners and non-owners. These include the presence of companions, trust in other travelers, and social pressures. We also identified differences in the factors that influence owners versus non-owners. For owners, additional considerations include whether the vehicle is shared with another regular driver and their perception of the accuracy of collected sensor data. For non-owners, additional factors include trust in stakeholders who are not traveling with them, such as the car owner or subsequent renters, and the duration of time spent in the vehicle.

This paper makes two main contributions. First, it explores car users' privacy concerns and identifies the key factors that influence the privacy preferences of both car owners and non-owners. Second, it provides a set of design suggestions to better address the diverse privacy needs of car users.

\section{Background and Literature Review}
We review prior research on privacy risks in connected vehicles, highlight studies exploring drivers’ privacy concerns and preferences, and discuss studies investigating users’ expectations for privacy in broader indoor smart environments.

\subsection{Privacy Risks in Connected Vehicles}
Modern vehicles are increasingly equipped with a wide range of sensors that enhance safety, personalization, and overall driving experience  \XDCMT{\cite{Sleeper,Halin}}. Some of these sensors are considered essential, such as Driver Drowsiness and Attention Warning (DDAW) systems \XDCMT{\cite{Palao}}. Importantly, many in-vehicle sensors are not limited to monitoring the vehicle’s mechanical or operational state. They also capture information about the activities and behaviors of users inside the cabin. Examples include microphones, cameras, millimeter-wave radars, electrocardiograms, pressure sensors, and optical sensors \XDCMT{\cite{Tang,Kafková,Nowara,A_Malik}}. These technologies enable vehicles to collect rich data about their occupants, including physiological and biological signals, and in some cases, even genetic information \XDCMT{\cite{Bellan_2023,Leonhardt}}. Consequently, modern cars can detect a wide array of user-related information beyond driving, ranging from daily routines and social interactions to health status and private behaviors \XDCMT{\cite{Crandall,Bellan_2023, András}}. 

While these capabilities offer opportunities for highly personalized experiences, they also raise important privacy considerations. Sensor data in vehicles is often insufficiently protected, creating vulnerabilities that can be exploited. Malicious actors may gain unauthorized access to user data through the vehicle’s connectivity features, implanting Trojans in vehicle buses or infotainment systems to capture sensitive information, such as location, travel itineraries, contacts, and call records \XDCMT{\cite{miller2015remote,Kim2}}. In some cases, attackers can even compromise physical vehicle interfaces to obtain driving trajectories or user identity information \XDCMT{\cite{Haohuang,Strandberg}}. Such breaches enable continuous tracking, extortion, blackmail, and even remote control of vehicle functions, amplifying users’ privacy concerns in modern automotive environments.

Over the years, several technical solutions have been proposed to address these challenges. At a system level, Sweeney, L. proposed to remove associations between data \XDCMT{\cite{Sweeney}}. Machanavajjhala et al. further proposed to eliminate the statistical significance of sensitive information on this basis to prevent homogeneity attacks and background knowledge attacks \XDCMT{\cite{Machanavajjhala}}. Additionally, the differential privacy-based risk quantification framework allows researchers to protect individual privacy by injecting carefully calibrated noise while ensuring the data remains usable \XDCMT{\cite{Dwork,Muhammad}}. Furthermore, researchers have proposed to utilize encryption protocols, periodic pseudonym certificates, and signatures during data transmission to prevent unauthorized access and continuous data tracking during transmission \XDCMT{\cite{Bhargavan,Khodaei,Brecht}}. While these approaches are important for protecting data at the technical level, they do not explain how users perceive, negotiate, and respond to privacy trade-offs in everyday cabin use.

\subsection{User Privacy Preferences for Vehicle Sensors}
Prior work has examined how people perceive privacy in relation to vehicle sensing technologies \XDCMT{\cite{Bella,Sleeper}}. Dowthwaite et al. studied the perspectives of car owners who primarily drive alone \XDCMT{\cite{Dowthwaite}}. Their findings revealed that many drivers view the cabin as a private space, which makes them particularly cautious about in-cabin sensing. Similarly, Walter et al. investigated drivers’ attitudes and found that the type of data collected by vehicle sensors significantly shaped their privacy preferences \XDCMT{\cite{Walter}}. Acharya et al. further showed that drivers’ willingness to share data depended on how their data would be used \XDCMT{\cite{Sailesh}}. Expanding beyond the driver’s perspective, Sleeper et al. explored privacy concerns from both drivers and non-drivers \XDCMT{\cite{Sleeper}}. Their study included passengers as well as individuals outside the vehicle, such as pedestrians and nearby drivers. This broader scope offered important insights into public safety considerations.

In car rental scenarios, prior studies have shown that renters often remain unaware that infotainment systems can retain their personal information \XDCMT{\cite{Carlton}}. To address this concern, researchers have developed tools that support renters in deleting their personal data once the rental period ends \XDCMT{\cite{Carlton}}. Beyond rental contexts, prior work has also shown that this preference for opting out of identifiable data collection extends beyond just car users. For example, Bloom et al. investigated pedestrians’ perceptions of privacy in relation to self-driving cars and found that more than half of participants (54\%) were willing to spend over five minutes in an online system to opt out of such data collection \cite{Bloom}. Despite these indications of concern regarding modern vehicle sensors, studies also suggest that users, such as drivers, are often open to sharing certain types of personal information when doing so leads to tangible benefits, such as enhanced safety \XDCMT{\cite{Dowthwaite}}. This finding resonates with the work of Bossauer et al., who examined car renters in car-sharing contexts and similarly observed a willingness to trade some degree of privacy for the practical advantage of vehicle access \XDCMT{\cite{Bossauer}}.

Our work differs from prior research on user preferences in sensor-enabled smart car cabins by adopting a broader perspective. While earlier studies have primarily examined privacy considerations in contexts where car owners drive alone, our study adopts a broader perspective to reflect how modern vehicle cabins are actually used. We examine how privacy preferences shift across a wider range of dynamic scenarios, particularly when the cabin becomes a socially shared space, such as when users travel with companions or switch between driver and passenger roles. Furthermore, we extend this work to non-owners, comparing their perspectives with those of owners to identify not only shared privacy concerns but also factors unique to each group. Building on this, we further explore how these privacy judgments are shaped by the relationships around multi-stakeholders, including the influence of trusted companions, absent car owners, employers, and even future users of the same vehicle.

\subsection{User Privacy Preferences in Indoor Smart Environments}
Within the broader ecosystem of smart environments, indoor settings have been widely examined with respect to privacy concerns and user privacy preferences. Prior research suggests that multiple factors shape these preferences and expectations. For example, the purpose of data collection, the context in which data is gathered, and the parties with whom the data is shared can all significantly influence privacy attitudes \cite{Zhang}. Importantly, privacy preferences often differ between space owners and visitors. Visitors, for instance, may feel less comfortable with high-fidelity sensing technologies \cite{Lee}, while owners may place greater trust in their own devices and the manufacturers behind them. Consequently, studies show that homeowners may not actively engage in privacy protection at home, which can lead to unintentional disclosures of personal information \cite{Yao}.

Even within the home, privacy preferences are not uniform. Research has demonstrated that location within the household plays a key role in shaping privacy perceptions and influencing the adoption of privacy settings \cite{Kim}. Visitors, in contrast, are frequently unaware that sensor data collection in smart environments may raise privacy concerns \cite{Yao, Marky}. Recent findings suggest that visitors seek greater transparency about why data is collected, how it is managed, and what security risks may be involved \cite{Marky}. In fact, visitors often express a need to evaluate the status of sensors in real time to better understand how their data is being captured \cite{Marky}.

Despite these needs, visitors frequently report feeling powerless in situations where sensor configurations are necessary to align with their privacy preferences. Social dynamics can further complicate this challenge. For instance, visitors may hesitate to adjust or disable sensors due to social pressures within the host’s home \cite{Emami}. To address such challenges, researchers have proposed the use of Privacy Assistants (PAs) to help users discover and manage smart environment sensors \cite{Das, Chow, Colnago}. For example, Fernandez et al.’s system \cite{Bermejo} enables users to view active sensors and monitor the data being collected. As demonstrated across several studies, PAs have the potential to improve user comfort and foster trust in unfamiliar smart sensing environments \cite{Emami, Hosub}.

\begin{table*}[!htb]
  \centering
  \caption{Demographic of Participants}
  \label{tab:demographic}
  \begin{tabular}{c c c c c l}
    \toprule
    Participant & Gender & Age & Driving Experience (years) & Starting Scenario &
    \makecell[l]{Countries Where Driving\\Experience Was Gained} \\
    \midrule
    P1  & M & 30-34 & 12 & Scenario 2 & China/Canada \\
    P2  & M & 30-34 & 16 & Scenario 1 & China/Canada/Zimbabwe \\
    P3  & M & 20-24 & 6  & Scenario 3 & China/Canada \\
    P4  & M & 20-24 & 6  & Scenario 3 & China/Canada \\
    P5  & M & 25-29 & 10 & Scenario 2 & China/Canada/USA \\
    P6  & M & 30-34 & 14 & Scenario 1 & China/Canada \\
    P7  & F & 30-34 & 12 & Scenario 1 & China/Canada \\
    P8  & F & 25-29 & 8  & Scenario 2 & China/Canada \\
    P9  & M & 20-24 & 5  & Scenario 3 & China/Canada/USA \\
    P10 & F & 20-24 & 5  & Scenario 2 & China/Canada/USA \\
    P11 & M & 20-24 & 6  & Scenario 3 & China/Canada/USA \\
    P12 & F & 25-29 & 9  & Scenario 1 & China/Canada \\
    P13 & M & 30-34 & 12 & Scenario 3 & China/Canada \\
    P14 & M & 20-24 & 6  & Scenario 1 & China/Canada \\
    P15 & M & 20-24 & 6  & Scenario 2 & China/Canada \\
    P16 & M & 25-29 & 9  & Scenario 1 & China/Canada \\
    P17 & F & 20-24 & 6  & Scenario 3 & China/Canada \\
    P18 & M & 25-29 & 10 & Scenario 2 & China/Canada \\
    \bottomrule
  \end{tabular}
\end{table*}

\section{Methodology}
To explore car users’ privacy preferences, we conducted one-on-one semi-structured interviews with 18 participants. Although other methods, such as focus groups, have been employed in similar studies \XDCMT{\cite{Yao,Colnago}}, we selected individual interviews to encourage participants to reflect independently on their experiences with in-car sensing systems, minimizing potential influence from others’ perspectives. Each interview lasted an average of 100 minutes, excluding breaks. The study was approved by our institution’s Institutional Review Board (IRB).

\subsection{Participants}
We recruited 18 participants (13 male, 5 female; mean age = 26.7 years, SD = 3.67, age range: 22-34) via word-of-mouth, social media, and university recruitment channels. The demographics of the participants could be found in Table \ref{tab:demographic}. To minimize priming effects and bias, recruitment materials described the study only as “an interview about your perceptions of in-car sensing systems” without any reference to privacy. All participants had at least five years of experience as drivers and/or passengers.

\subsection{Pilot Study} 
We conducted two rounds of pilot studies, each with two participants, to iteratively refine our interview questions and optimize the experimental protocol. These studies identified minor issues that informed design adjustments. For example, the initial study introduction and some interview questions inadvertently emphasized negative aspects of the sensors. To address this, we revised the presentation to be more neutral, providing balanced information on each sensor’s functionality, application scenarios, and potential privacy risks. This revision enabled participants to more thoroughly consider the trade-offs between functional benefits and privacy concerns.

\subsection{Study Procedure}
This study aimed to explore car users’ privacy preferences, examining differences between car owners and non-owners and identifying potential factors influencing their privacy perceptions of different in-car sensors. The study was structured into two phases.

\subsubsection{Phase 1: General Understanding.} 
Phase 1 was used primarily to establish participants’ baseline understanding and ground later scenario-based discussion. During this stage, participants were first invited to describe their prior understandings, direct experiences, and general perceptions of smart cabin sensors, including situations in which they had encountered sensing technologies as drivers, passengers, owners, or temporary users. We used this phase to elicit participants’ baseline assumptions before moving into scenario-based probing. We then presented a working definition: "Smart cabin sensors are sensors integrated into a vehicle’s interior to monitor the cabin environment, occupants, and their behaviors. These sensors use technologies such as infrared, ultrasonic, radar, and cameras to collect real-time user data, enhancing safety, comfort, and convenience."

To contextualize the discussion, we presented examples of state-of-the-art smart cabin sensors, including infrared and RGB cameras, microphones, 3D time-of-flight cameras, millimeter-wave radar, ultrasonic sensors, seat pressure and seatbelt sensors, steering wheel capacitive and steering angle sensors, and climate sensors such as temperature and air quality monitors. For each example, we described the type of data collected, its location within the cabin, its functional purpose, application scenarios, and potential privacy risks. We then introduced the concepts of car owners and non-owners. Car owners are people who purchase and own the car, whereas non-owners are those who neither own nor primarily use the car but may still be involved in its use, such as family members, friends, or renters. A detailed summary of the calibration procedure and the sensor categories introduced to participants is provided in the supplementary material.


\subsubsection{Phase 2: Scenario-Based Discussion.} 
Similar to \XDCMT{\cite{Yao, Colnago}}, we introduced three seeded scenarios to help participants better situate themselves in their roles and contextualize their privacy perceptions. The scenarios were inspired by common car usage contexts: (1) \textit{Personal car for multiple purposes}: You own a vehicle equipped with smart cabin sensors. You primarily use the car for family activities, such as driving your children to school or commuting to work, and occasionally for ride-hailing services such as Uber. The smart cabin sensors are active throughout daily use. (2) \textit{Rental car}: You rent a vehicle equipped with smart cabin sensors. While you are unsure of the exact sensors, you know they are always active. (3) \textit{Ride-hailing}: You take a ride-hailing service for a commute. The vehicle’s smart cabin sensors remain active throughout the trip. 

All participants were prompted using these same three seeded scenarios, which were designed to capture a diverse set of dimensions, including ownership (owner, non-owner), role (driver, passenger), cabin public status (semi-private, semi-public), and duration of car use (long-term: personal car, short-term: rental car; temporary: ride-hailing). Participants were not limited to these scenarios and were encouraged to discuss other contexts they deemed relevant. Additional scenarios introduced by participants were treated as extensions of these discussions rather than as separate analytic conditions, and were coded by mapping them onto the same dimensions. For each scenario, participants first reflected on the benefits and concerns of smart cabin sensors from the perspective of an owner, and then considered the same scenario from the perspective of a non-owner, again evaluating both advantages and potential issues. Participants demonstrated a good understanding of both the advantages and potential issues of smart cabin sensors, including privacy implications. We subsequently focused our discussion on participants' privacy concerns and preferences. More details about this phase can be found in the supplementary material.

\subsection{Data Analysis}
All interviews were audio-recorded with participants’ consent. Two co-authors transcribed the recordings, after which we collaboratively reviewed the first three transcripts to develop an initial codebook. We refined and finalized this codebook through multiple rounds of discussion. Using the initial version, we independently coded the remaining transcripts. After coding half of them, we met to calibrate our approach, incorporating any newly identified codes into the codebook. This process was repeated until all transcripts were coded. The final codebook yielded an inter-coder agreement (Cohen’s Kappa) of \BLCMT{0.83}, indicating good agreement \XDCMT{\cite{Fleiss}}. Further details of the final codebook are provided in the supplementary material.

Rather than coding each scenario as an isolated unit, we coded the full transcript and then identified factors recurring across contexts. Based on the final codebook, which included \BLCMT{99} unique codes (e.g., social pressure from travel partners, trust in travel partners), we conducted a cross-identity comparative analysis. We systematically grouped related codes into 9 categories to delineate themes with shared influence across both cohorts from those that disproportionately impacted either owners or non-owners. For instance, codes related to behavior changes, such as shielding personal smartphone screens or suppressing conversations in transit, were synthesized into Presence of Companion (C1, Sec.~\ref{Presence of Companion}). Similarly, trust-related codes, such as feeling secure with familiar co-passengers, were categorized into Trust in Present Companions (C2, Sec.~\ref{Perceived Trust in Travel Companions}).

\section{Results}
In this section, we present the findings of our study, organized around the key factors that shape the privacy preferences of car owners and non-owners.

\subsection{Participants’ General Perceptions}
\label{Participants’ General Perceptions}
During our discussions, participants described many benefits of integrating sensors into smart car cabins. Participants frequently described safety-related benefits as an important reason for accepting smart cabin sensing (P1-2, P6-9, P10, P11, P14, P17-18). Several participants also highlighted security and dispute related benefits, especially in ride-hailing, rental, or other commercial contexts where recordings could help retain evidence or protect drivers, passengers, and vehicles (P3, P4, P8, P10, P13, P15, P18). For example, some participants commended that sensors capable of tracking biometric indicators like driver fatigue could substantially enhance driving safety(P1, P3, P10, P11, P13). As P1 explains: “When it comes to life or property safety, I’m more tolerant”. Some participants also valued in-cabin recording or monitoring for security, accountability, and dispute resolution(P5, P18). Across all discussion scenarios, participants appreciated the value of these sensors, and several expressed enthusiasm about the increasing adoption of smart car cabins \BLCMT{(P9, P10, P18)}.

However, participants also expressed concerns about when sensor data could be used beyond its intended purpose, such as revealing personal characteristics, hobbies, or other information unrelated to the driving experience (P1, P4-12, P14-18). In general, participants were concerned about the potential for third-party access to sensor data, such as by automakers, data brokers, insurance companies, or other car users (P1, P4-7, P11, P14-16, P18). These privacy concerns varied depending on individual perspectives, specific scenarios, car ownership, and the user’s role in the vehicle (driver or passenger) (P1, P6, P7, P11, P14-16, P18). In this work, we focus on car ownership and examine the factors that influence the privacy preferences of both car owners and non-owners. 

Next, we first present factors that influence privacy preferences across both groups and then discuss factors that have a stronger influence on owners versus non-owners. Finally, we describe participants’ privacy-seeking behaviors and strategies for managing these concerns.

\subsection{Common Factors Influencing Privacy Preferences} \label{common factors}
We observed several factors that influence privacy preferences for both car owners and non-owners.

\subsubsection{Presence of Companion} (C1)
\hfill\\
\noindent
\label{Presence of Companion}
When driving alone, participants perceive the cabin as a semi-private space, less secluded than the home, but still comfortable enough to engage in activities they might avoid in more public settings \BLCMT{(P1, P3, P4, P5, P11, P12, P18)}. They recognized that the cabin is not entirely private, as windows allow others to see inside and voices can carry beyond the vehicle, particularly when parked on a busy street. As a result, participants often moderate their behavior, engaging in fewer private activities than they would at home (e.g., changing clothes). For example, P1 noted, “If the camera is on and I can’t turn it off, I wouldn’t feel comfortable taking a nap in my car during my lunch break, because I’d feel like someone was watching me. Nevertheless, the cabin provides a sense of freedom, supporting personal expressions, such as singing, activities that are uncommon in public settings. Within this context, users may also consider the implications of sensors that could capture sensitive data.

\begin{displayquote}
    \textit{“I don’t see the car cabin as completely private since people outside can still see me, but it feels kind of semi-private. Still, I wouldn’t want to be recorded if there’s a camera running in my cabin. Honestly, when I’m in a great mood and singing along, I really don’t want that recorded, especially because I’m a terrible singer.” (P1)}  
\end{displayquote}

Participants’ privacy preferences were influenced when they traveled with companions \BLCMT{(P1-18)}. For instance, some reported being more cautious about in-cabin sensors when driving with people they care about \XDCMT{(P1, P2, P9)}. This increased caution reflects a desire to protect not only their own privacy but also that of their passengers, particularly those who may not be fully aware of the sensors or the types of data being collected. Participants expressed concern that sensitive passenger information could be inadvertently shared during conversations. At the same time, there were situations in which participants relaxed their privacy requirements in the presence of passengers. For example, \XDCMT{P14} explained: 

\begin{displayquote}
    \textit{“I’m usually pretty cautious about riding in cars with cabin cameras because I find them invasive. But when I’m in my sister’s car with her baby, she has a baby-monitoring camera so she can keep an eye on the little one. I feel fine with that camera because I understand why it’s important for her.”}
\end{displayquote}

\subsubsection{Perceived Trust in Travel Companions} (C2)
\hfill\\
\noindent
\label{Perceived Trust in Travel Companions}
Perceived trust refers to car users' perceived level of trust toward others traveling with them. These companions may vary in their social relationship to the user (e.g., family members or strangers), in their connection to the vehicle (e.g., the car owner or a non-owner), or in their role in the car (e.g., driver or passenger). Several participants noted that their perception of privacy could be influenced by the level of trust they placed in others sharing the cabin \BLCMT{(P1, P2, P3, P6, P8, P11, P15, P16, P17, P18)}. For example, when traveling with less familiar people, such as strangers in carpooling or ride-hailing scenarios, participants tend to perceive the cabin as a semi-public space and feel more comfortable with the presence of sensors than when traveling with trusted companions or alone \XDCMT{(P1, P6, P16, P18)}. This change seems largely driven by accountability concerns, which are more prominent in these contexts. Participants reported that in such situations, the car is no longer seen as a semi-private space, and they naturally adjust their behavior to avoid revealing sensitive personal information. In this context, sensors serve as a reliable tool to manage potential conflicts with passengers. As P18 explained:

\begin{displayquote}
    \textit{"A ride-hailing passenger once asked me to make a U-turn over a solid line, saying everyone does it around here, to take a shortcut because they were running late. I refused, and they ended up filing a complaint since we were a bit late. Luckily, I had an audio recording, and once I showed it to the company, they didn’t penalize me. Without that sensor, I wouldn’t have been off the hook." }
\end{displayquote}

\subsubsection{Social Pressure} (C3)
\hfill\\
\noindent

Social pressure refers to the influence car users experience from the people they travel with. Several participants discussed that the presence of sensors in vehicles can create subtle social pressures, especially when passengers have differing privacy preferences \BLCMT{(P3, P4, P11)}. Some people may be skeptical of the technology, while others are less concerned. Participants described moments of discomfort when riding in sensor-rich cabins but chose to proceed nonetheless, influenced by the preferences of their companions \BLCMT{(P8, P15, P17)}. Others reported that such differences sometimes led them to avoid vehicles equipped with sensors, such as cameras \BLCMT{(P6, P9, P16)}. For instance, \XDCMT{(P9)} described their partner was highly concerned about privacy in smart environments led them to opt out of riding in electric cars equipped with extensive sensing capabilities. Participants also told us about the frustration they had when they had to explain and justify the privacy implications of these sensing systems to their passengers, often even when the participants themselves were unsure about the risks \XDCMT{(P5, P13, P14)}. 

\begin{displayquote}
    \textit{“There was this one time I picked up someone I didn’t know very well, and he noticed the cabin camera and asked ‘Why is there a camera here? Are you monitoring me?’ I’m not sure if he was completely serious, but I felt a little awkward. I mean, I didn’t put it there, it came with the car. Honestly, I wasn’t even sure if it was monitoring us, but I still tried to reassure him that it wasn’t a big deal. I hoped I was right. Still, I have to admit, I really wished the camera hadn’t been there at that moment.” \BLCMT{(P14)}}
\end{displayquote}

Certain safety-related sensors, such as fatigue detection, alert drivers when potentially unsafe situations are detected. Passengers may interpret these notifications in ways that unintentionally create additional pressure on the driver. For example, a mandatory audible alert from a fatigue detection system could be perceived by passengers as signaling that the driver is distracted, overly tired, or lacking skill. Although such alerts generally reflect important safety considerations, occasional false detections may occur. In these cases, car drivers may feel frustration toward the sensors, even as they acknowledge their overall value for enhancing safety.

\begin{displayquote}
    \textit{"When I’m driving with someone else in the car, the sensor alerts go to everyone, not just me. Most of the time, they don’t really know what’s happening with me or on the road, but they still feel like they can judge me." (P3)} 
\end{displayquote}

\subsubsection{Data Type} (C4)
\hfill\\
\noindent
Participants expressed varying privacy preferences depending on the type of data collected by smart cabin sensors. For instance, they were more cautious about visual and auditory data from cameras and microphones, while showing comparatively less concern for information such as air quality, seat pressure, or physiological signals \XDCMT{(P1, P4, P7, P8, P11, P13)}. People may weigh the importance of different data types to their privacy in different ways. For example, some participants viewed data such as driving behavior positively, especially when it could lead to clear benefits, such as cost savings or increased safety \XDCMT{(P10, P11, P13)}. Others, however, disliked the idea, particularly when the same data was collected in vehicles they didn’t own. They worried this information might be accessible to third parties and therefore felt more concerned \BLCMT{(P5, P15, P16)}. In those cases, participants preferred that the data not be collected at all. As \XDCMT{P5} put it:

\begin{displayquote}
    \textit{“I used to drive for a ride-hailing company, and I rented the car from them. The cabin has a camera, and I get why recording video can help resolve conflicts on the job. But I also feel that some types of data collection go too far for a company car, even during work hours. For instance, I’ve heard that some vehicles can capture biometric data. I’m not sure if my company’s car does that, but if it does, I’d be uncomfortable using it, even for work.”}
\end{displayquote}

\subsubsection{Cost-Benefit Trade-off} (C5)
\hfill\\
\noindent
Although participants expressed concerns about privacy related to in-cabin sensors, they also acknowledged scenarios where these sensors could offer meaningful benefits. In some cases, participants were willing to give up a degree of personal privacy for these advantages (P10, P11, P13). For example, as noted earlier, some participants were open to sharing their driving behavior data with insurance companies if it could result in lower premiums, showing how the potential benefits can influence privacy considerations. Additionally, several participants indicated a willingness to share cabin data if it could assist in managing traffic incidents (\XDCMT{P10}). This aligns with findings from previous research \XDCMT{\cite{Dowthwaite, Bossauer}}.

\begin{displayquote}
    \textit{“My car insurance company collects my driving data to determine my premium. I understand there are privacy concerns, but I’m not worried because I consider myself a responsible driver. I also carpool with colleagues, and they know I drive safely. I feel that the sensors collecting my driving data are like my colleagues vouching to my insurance company that I am a good driver, which is helpful.” (P10)}
\end{displayquote}

\subsection{Factors with Stronger Influence on Car Owners}
In addition to the factors that could commonly influence privacy preferences for both car owners and non-owners, we observed factors that could have a stronger influence on the car users, who own the vehicle.

\subsubsection{Single-User or Shared Vehicle} (O1)
\hfill\\
\noindent
Single-user scenarios refer to situations in which the car owner is the sole driver of the vehicle. In contrast, shared-use scenarios occur when the owner allows others, such as family members, siblings, or roommates, to drive the car regularly. Several participants discussed this scenario \BLCMT{(P2, P3, P10, P12, P14)}. For instance, \BLCMT{P14} explained that they typically drove the car during the day for school, while another household member used it at night for work. Participants in shared-use settings often raised different privacy concerns compared to the car owners who primarily drove alone. In particular, the car owners worried that additional drivers might gain access to sensor data. As a result, participants who shared their car tended to be especially cautious about how smart cabin sensors, such as cameras and microphones, might capture sensitive personal information. In contrast, car owners who primarily drove their vehicles themselves did not express similar concerns. 

\begin{displayquote}
    \textit{“I share my car with my roommate, who also helps pay off the loan. Financially, it works out great, but sometimes I feel uneasy about whether the camera or microphone data could be accessible to him. I don’t know if that actually happens, but just the possibility makes me worried. I really value having private moments in my car that I don’t want anyone else to know about, and honestly, I don’t want to know about his private moments either. In that sense, the sensors sometimes feel a little too intrusive for us.” \XDCMT{(P12)} }
\end{displayquote}

\subsubsection{Perceived Data Accuracy} (O2)
\hfill\\
\noindent
Perceived accuracy of sensor data refers to car owners’ perceptions of how precisely sensors capture information related to their privacy. While many participants had limited awareness of this issue, some \BLCMT{P6} noted that even without highly intrusive sensors like cameras, combining data from various sources, such as seat pressure or biometric sensors, could potentially reveal a comprehensive picture of their characteristics, habits, preferences, and daily routines. This perception of accuracy could influence car owners' privacy preferences. For instance, \BLCMT{P14} indicated that they would be more concerned about cabin sensors if the information gathered could reliably reflect private details unrelated to driving. 

Perceived accuracy is not necessarily influenced by sensor performance alone. It can also depend on car usage patterns, such as whether the vehicle is shared among multiple drivers. When cars are shared among multiple drivers, sensor data may contain noise from different users. In these cases, participants mentioned that they would feel less concerned about privacy risks, as the additional noise could create a sense of protection \BLCMT{(P2, P3, P12, P13)}. In contrast, when a car is driven exclusively by its owner, sensor data reflects mainly their own information, creating the perception that more detailed personal information might be inferred. For example, \BLCMT{P10} remarked:

\begin{displayquote}
    \textit{"I just stopped sharing my car with my roommate last month. Now I guess whoever steals my data will know me much better, since it’s only my data. If the air-quality sensor picked up smoke, they could easily figure out that I smoke, especially because no one else drives my car."}
\end{displayquote}

\subsection{Factors with Stronger Influence on Non-Owners}
We also observed several factors that influence the privacy preferences of temporary car users, who do not own the vehicles, more strongly than those of car owners.

\subsubsection{Perceived Trust in Absent Stakeholders} (NO1)
\hfill\\
\noindent
Absent stakeholders are those who are not physically present in the vehicle (in contrast to cases where stakeholders are inside the car, as discussed in Section \ref{common factors}). Examples include the future users of the car or the car owner when they are not traveling with the current user. When the absent stakeholder is the car's owner, non-owners’ privacy preferences can vary depending on their level of trust in that owner. For instance, participants described that when driving a family member’s car, they tend to be less concerned about data collection \XDCMT{(P7, P11, P14, P15)}. They explained this by reasoning that trusted people are less likely to share their data with others, or by assuming they would have easy access to and control over the data themselves. Note, these assumptions may not always reflect reality, as even the car owner may have limited control over the data collected by the vehicle.

In contrast, participants expressed stronger privacy concerns when driving a car owned by someone they did not fully trust \XDCMT{(P2, P10, P14)}. For example, in work-related contexts, participants who regularly used company cars tended to be more cautious about the types of data sensors that might capture and potentially disclose to their employer. As \XDCMT{(P8)} explained.

\begin{displayquote}
    \textit{“I drive a company car every day to different worksites. I usually avoid doing anything personal during work hours, but on lunch breaks, I often end up eating in the car if there’s nowhere else to go. That’s when I’ll make personal calls or even sing along to the radio, things I definitely wouldn’t want my company to know about.” }
\end{displayquote}

Participants also raised privacy concerns when they lacked trust in subsequent users of a shared vehicle, such as renters or coworkers \BLCMT{(P4, P11, P16)}. They worried that the “sensor footprint” they left behind might be visible not only to the car’s owner or manufacturer but also to subsequent users. These concerns were especially pronounced when participants did not know who would use the vehicle next. When the future driver was a familiar person, such as a family member, participants felt somewhat more at ease, but their concerns did not completely vanish.

\begin{displayquote}
    \textit{“I’m not sure if the next driver can access the sensor traces I leave behind in a rental car. For example, whenever I rent one, I often see other people’s phone IDs still saved in the Bluetooth settings. Sometimes the IDs even reveal their identity if they used their real name. This makes me wonder whether similar traces might be left through other sensors in the car as well.” \XDCMT{(P4)}}
\end{displayquote}

\subsubsection{Length of Car Possession} (NO2)
\hfill\\
\noindent
Participants’ privacy concerns varied depending on the duration of time they spent in a car they did not own, whether as a driver or a passenger. Some reported that during short trips, they were less concerned about the presence of sensors or the types of data being collected \BLCMT{(P4, P7, P11)}. On longer trips, however, their concerns grew. Some explained that, on short rides, they could easily adjust their behavior, such as limiting phone calls, to reduce the exposure of sensitive information, but maintaining these adjustments over longer trips was more challenging \BLCMT{(P4, P14, P15)}. Others noted that longer trips naturally involve more extensive data collection, which they perceived as increasing potential privacy risks \BLCMT{(P16, 17)}, although the actual amount of sensitive information potentially exposed may not directly correlate with trip length \XDCMT{\cite{atockar_2014,Xun}}. 

\begin{displayquote}
    \textit{"I feel like more of my personal data can be collected in rental or work vehicles than in a cab I took the other day, because the car has more time to collect my data." \BLCMT{(P16)}}
\end{displayquote}

While our findings do not define a universal threshold for what constitutes a “short” versus “long” trip, participants frequently described city commutes as short, whereas rental cars were often considered long-term use, since rentals typically last a full day or more \BLCMT{(P4, P7, P11, P14, P15)}.

\subsection{Privacy Seeking Behaviors}
\label{Privacy Seeking Behaviors}
Participants described strategies they currently use, or could envision using, to protect their privacy in cars. Many noted awareness that some sensors can be disabled at the software level, though both owners and non-owners were often hesitant to do so due to uncertainty about the consequences. Disabling some sensors could unintentionally affect other features, and the impact of these changes was often unclear. Sensors can also be disabled at the hardware level. However, a key challenge was the limited understanding of which sensors are present in their vehicles, where they are located, and how to access them. For example, most participants were unsure about the placement or even the number of microphones in their cars. When sensors were more visible or accessible, such as cameras, some participants reported occasionally taking steps to disable them. One participant \BLCMT{(P16)} shared that they purchased a cover for their electric car’s in-cabin camera, allowing them to block it whenever desired. 

For most participants, however, privacy management rarely went beyond such measures. They noted that disabling sensors often requires technical skills or specialized tools, and once a sensor is disabled, restoring its functionality can be challenging. One participant \BLCMT{(P16)} mentioned that some automakers provide service manuals showing sensor locations and removal procedures, and there are even video tutorials available online. Despite this, most participants felt these steps were too complex for the average car owner, leaving them with little choice but to keep sensors active. 

Some participants were aware of alternative approaches, such as using devices to interfere with sensors. For example, one participant \BLCMT{(P14)} described seeing a product on Amazon that emits ultrasonic noise to disrupt car microphones and prevent conversations from being detected. Others \BLCMT{(P8)} reported that they had intentionally chosen to avoid purchasing or riding in cars equipped with cabin sensors as a way to protect their privacy.

Participants also adopted strategies to temporarily manage their privacy concerns without directly interacting with the sensors. For instance, some reported wearing a face mask while riding in ride-hailing vehicles equipped with cabin cameras \BLCMT{(P7, P14, P15, P16)}. Others were mindful of modifying their behavior, such as avoiding discussions of sensitive information during the ride \BLCMT{(P7, P8, P18)}. One participant \BLCMT{(P12)} described intentionally positioning their phone screen away from the cabin camera to prevent personal information from being visible.

\section{Discussion}
We will first summarize the new insights uncovered in our study in comparison to prior research. We will then reflect on the similarities and differences in privacy preferences between car users and users of smart indoor environments. Next, we will discuss the user perception gaps revealed by our findings. Finally, we will present several design recommendations for future privacy-enhancing mechanisms for smart car cabins.

\subsection{Summary of New Insights}
Previous research on privacy in connected cars has primarily focused on scenarios where car owners drive alone \XDCMT{\cite{Dowthwaite, Sleeper}}. These studies suggest that owners often perceive their car as a private space and are willing to trade certain aspects of privacy for tangible benefits, such as improved safety \XDCMT{\cite{Dowthwaite, Bossauer}}. Other work has highlighted that owners’ privacy concerns also vary depending on the type of data collected by the vehicle \XDCMT{\cite{Walter}}. Building on this foundation, we explore the factors that influence car users' privacy preferences across a broader range of car usage scenarios, including situations where owners travel alone or with others, act as drivers or passengers, and when non-owners use the vehicle under similar situations.

Consistent with previous findings, we observe that factors such as data type and cost-benefit trade-offs significantly shape privacy considerations (C4, C5). This suggests that participants approached privacy decision-making from a pragmatic perspective, weighing different forms of data collection against the specific safety, accountability, convenience, or economic benefits they perceived \cite{Westin1967PrivacyFreedom}. Beyond these, by examining cases where the car functions as a semi-public space, we identify additional influences on privacy perceptions for both owners and non-owners, including the presence of companions, trust in travel companions, and social pressures (C1, C2, C3). Our findings further reveal distinctions in influential factors between owners and non-owners. For owners, additional considerations include whether the vehicle is shared with another regular driver and the owners' perceived accuracy of the collected sensor data (O1, O2). For non-owners, additional factors involve trust in stakeholders not traveling with them, such as car owners or subsequent renters, and the duration of their time in the vehicle (NO1, NO2). Compared to prior research, these insights provide a richer understanding of car users’ privacy preferences and inform more precise design implications for developing effective privacy-protection mechanisms in smart cabins (discussed later in this section).

\subsection{Smart Car Cabin vs Smart Indoor Environments}
Cars and indoor environments, such as homes, are important components of the broader ecosystem of smart environments. Previous research has shown privacy concerns in smart homes \cite{Choe,Emami,Zhang,Marky,Yao} as well as in cars \cite{Dowthwaite, Sleeper, Walter}. Our study builds on this work and shows that, although privacy issues arise in both settings, people tend to think about them in different ways. This means that lessons learned from smart home privacy cannot directly be applied to cars. We unpack several key differences that influence how privacy is experienced in smart homes compared to smart car cabins.

\subsubsection{Awareness, Transparency, and Control}
\hfill\\
\noindent
In smart homes, owners are often aware of sensors because they play an active role in installing or configuring them \XDCMT{\cite{Yao}}. In contrast, such awareness does not always extend to car owners when purchasing a vehicle. As several of our participants \BLCMT{(P1, P5, P6, P14)} observed, car manufacturers and dealerships tend to emphasize features that are powered by sensors, rather than explaining the underlying technologies, the types of data being collected, or the related privacy implications (see sec. \ref{Participants’ General Perceptions}). For instance, fatigue detection is frequently promoted as a safety feature, yet participants noted that they were seldom informed about the specific sensors, such as cameras or heart rate monitors, that make this capability possible. 

Furthermore, sensors in cars are often much less visible than those in smart homes. While many home sensors are added through retrofitting and can be more noticeable, smart cabin sensors are seamlessly integrated into the interior. Consequently, our findings indicate that non-car owners are generally unaware of the sensors embedded in vehicles, echoing research from smart homes, where visitors are often unfamiliar with the sensors present \XDCMT{\cite{Yao,Marky}}. This lack of awareness also extends to the vehicle owners themselves.

Prior research on smart homes indicates that both owners and bystanders, though somewhat limited in addressing their privacy concerns, often retain a degree of control over sensors \XDCMT{\cite{Marky,Yao}}. In contrast, participants in our study described controlling smart cabin sensors as particularly challenging \BLCMT{(P3, P11, P18)}. For instance, \BLCMT{(P18)} noted that managing or configuring in-cabin sensors was difficult, not only because the controls were hard to find, but also because disabling or reconfiguring a sensor could unintentionally affect multiple other vehicle functions (See sec \ref{Privacy Seeking Behaviors}).

\subsubsection{Factors Influencing Privacy Preferences}
\hfill\\
\noindent
Prior research in indoor environments suggests that users’ privacy preferences and expectations are shaped by factors such as the purpose of sensor data collection and the parties with whom data is shared  \XDCMT{\cite{Zhang,Emami}}. Our study with car users indicates that similar patterns also appear in cars. In smart home studies, non-owners’ privacy perceptions have been shown to depend on factors such as perceived device utility, social relationships, trust, and length of stay \XDCMT{\cite{Yao}}. At a high level, several of these factors align with our findings in the automotive context, such as perceived trust and duration of use. For example, we observed that non-owners’ trust in other stakeholders, such as the car owner, plays a significant role in shaping their privacy preferences in smart car cabins (NO1). However, the influence of these factors manifests differently in car settings. While trust is a central factor in both domains, vehicle users also consider additional stakeholders, including future occupants of the same car, which introduces further complexity (NO1). Moreover, the presence of companions, who may or may not be familiar to the user, adds another dimension to privacy considerations (C1, C2). Taken together, these findings highlight the unique nature of cars as dynamic environments where mobility and social interactions shape privacy expectations in ways that differ from homes or other indoor settings \cite{Altman1975Environment}. 

\subsubsection{Privacy Seeking Behaviors} 
\hfill\\
\noindent
We observed similarities and differences in how users manage their privacy across indoor environments and smart car cabins. For instance, users in smart indoor spaces sometimes choose to avoid areas equipped with sensors \XDCMT{\cite{Yao}}. Similarly, in the context of vehicles, our participants reported choosing not to ride in cars with cabin sensors as a precaution to protect their privacy. In smart homes, visitors sometimes temporarily disable sensors that raise privacy concerns, for example, by covering a camera with a cloth. In contrast, in cars, such actions are more commonly observed among owners, who may use commercially available covers for cabin cameras (Sec \ref{Privacy Seeking Behaviors}). This difference may reflect social pressures on non-owners, who might feel hesitant to intervene, or the possibility that disabling sensors in cars could affect other vehicle functions (C3, Sec \ref{Privacy Seeking Behaviors}). Additionally, the potential consequences of disabling sensors are often unclear, so both owners and non-owners expressed hesitation about doing so. Another difference is that non-owners of cars often rely on short-term strategies to protect their privacy during brief trips. For instance, they may cover their face, avoid sensitive conversations, or postpone phone calls until after the ride (NO2). These behaviors are less commonly reported in indoor contexts, likely because longer stays make temporary measures less practical. 

\subsection{User Perception Gap}
Our study results reveal a gap between users’ perceptions and the actual state of sensor data management, which could influence privacy-related decisions. 

\subsubsection{Perceived Data Ownership}
\hfill\\
\noindent
Through our interviews, we found that participants often expressed a strong desire to maintain ownership of the data captured by in-cabin sensors when driving their own cars. This sense of ownership was especially clear when participants drove alone, as the data was perceived to reflect only their personal activity. Our findings resonate with prior research based on psychological ownership \XDCMT{\cite{Jon_L1,Jon_L}}, which suggests that car owners often feel a stronger responsibility to manage and control sensor data from connected vehicles \XDCMT{\cite{Cichy}}. 

Interestingly, this sense of ownership often extended beyond their own driving data. Even when passengers were present, many participants believed that they still “owned” the data generated by others in the vehicle. This was reflected in their perceived responsibility to protect passengers’ privacy. Notably, this perception held true regardless of whether the car owner was driving or riding as a passenger. Participants described feeling accountable for safeguarding the personal information of others, particularly when passengers were close friends or family members. In such situations, some car owners reported even greater privacy concerns, as they sought to protect both their own and their passengers’ sensitive data. In contrast, the sense of ownership was noticeably weaker when participants drove vehicles that did not belong to them, even though the sensors were collecting their personal data. As reflected in non-owners’ perceptions of trust toward the vehicle’s owner, many participants (e.g., P7, P11, P14, P15) expressed that they felt their data ultimately resided in the hands of the car owner. They also noted that it was up to the owner to determine how the data might be used, and that they should seek permission from the owner to access or control it. These perspectives illustrate that non-owners often do not perceive strong ownership over their sensor data. From both owners’ and non-owners’ viewpoints, the sense of data ownership appeared to be strongly tied to car ownership itself. One possible explanation for this perception is that non-owners typically have very limited access to their own data. This highlights an opportunity to design mechanisms that empower non-owners with greater access to and control over their personal data, as well as the information flow under different contexts \cite{Nissenbaum2004ContextualIntegrity}.

\subsubsection{Perceived Control Over Data}
\hfill\\
\noindent
Perceived control over sensor data refers to the extent to which car users believe they can access, manage, and determine how their information is shared with third parties. Our findings indicate that non-owners often felt limited in this regard, as they did not perceive ownership of the data collected by smart cabin sensors. In contrast, some owners described a stronger sense of control when voluntarily sharing their activity data with automakers to receive personalized services. They emphasized that the ability to opt in or withdraw at any time gave them confidence in managing their data. Participants further noted that this autonomy reduced their perception of privacy risks, making them more willing to share data when clear and tangible benefits were offered. This observation resonates with prior work in other domains, where transparent and accessible control mechanisms similarly encouraged data sharing \XDCMT{\cite{Xusen,Cichy,Brandimarte}}. However, not all car users were aware of other stakeholders who may also have control over sensor data, such as car manufacturers. Several participants expressed surprise that car owners were not the only ones with access to this data \BLCMT{(P4, P11, P14)}. Many were unaware that manufacturers could, for example, sell data to third parties, such as data brokers, without users’ awareness.

\subsubsection{Perceived Trust Over Stakeholders}
\hfill\\
\noindent
Our findings indicate that car users’ trust in different stakeholders strongly influences their privacy preferences within vehicles. When trust is appropriately placed, it enables users to make more informed decisions about in-cabin sensors, for example, being comfortable with cabin cameras while providing ride-hailing services to strangers. In contrast, trust in stakeholders who lack full control over data, or when users are unaware that control of their data is shared with other parties, can reduce caution and increase the risk of exposing sensitive information. Some participants reported being less careful about private activities when using a vehicle owned by a trusted household member. In these situations, private information could potentially be accessed by other stakeholders, such as automakers, without the user’s knowledge.

\subsubsection{User Awareness of Privacy Risk-Benefit Trade-offs}
\hfill\\
\noindent
Our findings indicate that both owners and non-owners encounter situations where they must negotiate the trade-off between maintaining privacy and relinquishing it for perceived benefits. A common motivation for such trade-offs is the promise of advantages, such as personalized in-cabin experiences or reduced insurance premiums, aligning with prior findings in automotive contexts \XDCMT{\cite{Bossauer,Dowthwaite}}. We also observed that social pressure influences privacy-related decisions. For example, people who typically avoid cars equipped with cabin sensors sometimes chose to ride in them when their companions expressed a preference for such vehicles. 

From our results, we observed three issues related to these trade-offs. First, the consequences of giving up privacy are not always clear. Several participants reported subscribing to telematics and connected vehicle services. Yet, many did not realize that their data could also be accessed by third parties, such as insurance providers, to train machine learning models that could affect their premiums \XDCMT{\cite{James}}. In such cases, participants believed they were exchanging privacy for improved driving experiences, while remaining unaware of broader implications, such as adjustments to insurance rates \BLCMT{(P6, P13, P17)}. Second, participants expressed uncertainty about the actual value of their personal data \BLCMT{(P2, P5, P13, P14)}. For example, some did not realize that activity data shared with telematics service providers could be sold to other companies without their awareness or compensation \XDCMT{\cite{DAD_Nguyen_2025}}. Third, our results indicate that privacy preferences vary depending on the type of data collected. Many participants regarded cameras and microphones as particularly privacy-sensitive and were therefore hesitant to share such information. By contrast, they were more willing to disclose data they perceived as less sensitive, such as physiological signals (e.g., heart rate, breathing patterns, and other biometrics) \BLCMT{(P8, P13, P14)}. However, most participants were unaware that these signals can also reveal deeply personal information, including health conditions, lifestyle habits, and even cognitive traits \XDCMT{\cite{Hidalgo,Avram,Martin}}.

\subsection{Design Implications}
Based on our study results, we make the following design suggestions to guide the development of future privacy-enhancing mechanisms in smart car cabins.

\subsubsection{Ensure Transparency and Control}
\hfill\\
\noindent
The concept is not new within the broader field of privacy design for smart environments. However, the most critical design requirement for sensor-powered car cabins remains the assurance of transparency and control, which is still largely missing (C2, C3, C4, C5, NO1, NO2). First, transparency should be guaranteed regarding the presence and functionality of sensors for all users. This responsibility begins with car manufacturers, who should clearly communicate the types and purposes of embedded sensors to car users. Car owners should also make this information easily available to other users of the vehicle. Additional stakeholders should also contribute to making this information accessible to potential users. For instance, in ride-sharing contexts, a promising strategy is to integrate transparency features directly into ride-sharing applications, enabling passengers to understand potential privacy implications before entering a vehicle or being exposed to sensors. Second, both drivers and passengers should have straightforward mechanisms to control sensor operations. Our findings show that many car users worry that disabling a sensor might compromise the functioning of other features. If such dependencies cannot be avoided from a technical standpoint, these consequences should be communicated clearly, along with alternative approaches that can provide a comparable level of privacy protection.

\subsubsection{Customizable Sensor Configurations to Support Diverse Privacy Needs}
\hfill\\
\noindent
Car travel often involves companions, which introduces dynamic and evolving privacy needs. Addressing these needs requires flexible, on-the-fly adjustments to in-car sensor configurations to maintain appropriate levels of privacy (C1, C2). For example, cabin cameras and microphones may be enabled when a ride-sharing driver picks up a passenger and disabled once the passenger leaves. To support such scenarios, privacy protection systems should allow car users to easily personalize sensor configurations to fit their preferences. Providing shortcuts or presets can further streamline this process, enabling users to switch quickly between different privacy settings.

\subsubsection{Enable Personalized Privacy Modes for Shared Car Use} 
\hfill\\
\noindent
Privacy protection systems should also account for common shared-use scenarios, such as work vehicles, rental cars (NO1), or cases where owners regularly permit others to drive their car (O1). Drawing inspiration from guest modes on personal computers, in-car privacy systems could provide configurable modes linked to user accounts. Each mode could offer tailored sensor settings to reflect diverse privacy preferences and usage contexts. In such systems, shared drivers could log into the vehicle’s computing interface to activate their preferred mode, ensuring that privacy protections are aligned with their individual needs. Additionally, each mode should clearly indicate which user’s data is being collected, helping all users understand the privacy implications of shared car usage (O2).

\subsubsection{Enable Car Users To Access Their Own Data}
\hfill\\
\noindent
All car users should have access to the data collected by cabin sensors, regardless of vehicle ownership. Privacy mechanisms should be designed to allow non-owners to independently view and manage their data without needing permission or assistance from the car owner. Such access could be provided through online protocols or mobile applications. For instance, in ride-sharing scenarios, passengers’ data could be made accessible via the ride-sharing app, enabling them to manage their sensor data even without physical access to the vehicle. This is particularly important when passengers no longer have access to the car or may feel social pressure not to request data from the owner.

\subsubsection{Clarifying the Cost of Sharing Privacy Data}
\hfill\\
\noindent
Car users, whether they own the vehicle or not, often encounter situations where they must weigh the benefits of sharing personal data against potential risks to privacy or the pressures of social expectations. To support informed decision-making, privacy mechanisms in smart car cabins should present these trade-offs clearly and transparently (C5). For instance, systems could notify users about how their data might be accessed, monetized, or shared with third parties, while also clarifying how such data might be used in ways that are not immediately visible (C4). Furthermore, these mechanisms should draw attention to how various types of sensor data, even those perceived as less sensitive than cameras or microphones, can still reveal significant personal information (O2, NO2). Additionally, making this information available to people who might exert social pressure could potentially help reduce undue influence and support more informed, mutually agreeable decisions (C3).


\section{Limitations and Future Work}
Our study has several limitations that should be acknowledged. First, our findings should be interpreted as exploratory and contextually situated. Most participants had car-use experience in Asia and North America, and these backgrounds may have shaped their expectations around surveillance, safety, and data practices. Future work with larger and more locally grounded samples could test the robustness of these themes across different cultural and usage contexts.

Second, given the exploratory focus on qualitative insights from car users, we did not engage in designing privacy protection systems for smart cabin sensors. Future research could leverage participatory design workshops to gain a deeper understanding of users’ expectations and requirements for such systems. Additionally, adopting mixed-method approaches, including controlled experiments, could help quantify the prevalence of these preferences and further validate and extend the findings of this study.

Finally, while our study offers several design implications, implementing these recommendations poses technical challenges. For instance, supporting customized control and configuration of cabin sensors, along with providing secure access to sensor data for multiple stakeholders, requires thoughtful planning and engineering. These challenges, however, create exciting opportunities for interdisciplinary collaboration among HCI researchers, engineers, and the automotive industry, paving the way for more user-centered smart cabin technologies.

\section{Conclusion}
In this paper, we investigate users’ privacy preferences in sensor-powered smart car cabins. While prior research has primarily focused on scenarios in which vehicle owners drive alone, we extend this exploration to include two common stakeholder groups: vehicle owners (i.e., individuals who purchase and own the car) and non-owners (i.e., individuals who temporarily use the car, such as family members, friends, or renters). Our study considers more realistic usage scenarios, including situations where owners travel alone or with others, both as drivers and passengers, as well as cases where non-owners occupy the vehicle under similar circumstances. Through semi-structured interviews with 18 regular car users, we found that, in addition to previously reported factors such as data type and cost–benefit trade-offs \XDCMT{\cite{Walter,Dowthwaite}}, other considerations strongly influence privacy preferences, including the presence of companions, trust in fellow travelers, and social pressures. Moreover, we observed differences between owners and non-owners in the factors shaping their privacy decisions. For owners, key considerations include whether the vehicle is shared with another regular driver and their perception of the accuracy of collected sensor data. For non-owners, important factors include trust in stakeholders not currently traveling with them—such as the car owner or subsequent renters—and the duration of time spent in the vehicle. Our findings provide valuable insights for designing user-centered smart car cabins that more effectively address users’ privacy needs.

\bibliographystyle{ACM-Reference-Format}
\bibliography{reference}

@article{Mishra,
author = {Mishra, Ashutosh and Lee, Sangho and Kim, Dohyun and Kim, Shiho},
year = {2022},
month = {06},
pages = {4360},
title = {In-Cabin Monitoring System for Autonomous Vehicles},
volume = {22},
journal = {Sensors},
doi = {10.3390/s22124360}
}

@article{Avram,
author = {Avram, Robert and Tison, Geoffrey and Kuhar, Peter and Marcus, Gregory and Pletcher, Mark and Olgin, Jeffrey and Aschbacher, Kirstin},
year = {2019},
month = {03},
pages = {16},
title = {PREDICTING DIABETES FROM PHOTOPLETHYSMOGRAPHY USING DEEP LEARNING},
volume = {73},
journal = {Journal of the American College of Cardiology},
doi = {10.1016/S0735-1097(19)33778-7}
}

@article{Lu,
author = {Lu, Jiayi and Peng, Zhaoxia and Yang, Shichun and Ma, Yuan and Wang, Rui and Pang, Zhaowen and Feng, Xinjie and Chen, Yuyi and Cao, Yaoguang},
title = {A review of sensory interactions between autonomous vehicles and drivers},
year = {2023},
issue_date = {Aug 2023},
publisher = {Elsevier North-Holland, Inc.},
address = {USA},
volume = {141},
number = {C},
issn = {1383-7621},
url = {https://doi-org.proxy.lib.sfu.ca/10.1016/j.sysarc.2023.102932},
doi = {10.1016/j.sysarc.2023.102932},
journal = {J. Syst. Archit.},
month = aug,
numpages = {15}
}

@ARTICLE{Gong,
  author={Gong, Zheng and Yang, Xuezhi and Song, Rencheng and Han, Xuesong and Ren, Chong and Shi, Hailin and Niu, Jianwei and Li, Wei},
  journal={IEEE Transactions on Instrumentation and Measurement}, 
  title={Heart Rate Estimation in Driver Monitoring System Using Quality-Guided Spectrum Peak Screening}, 
  year={2024},
  volume={73},
  number={},
  pages={1-14},
  doi={10.1109/TIM.2024.3352710}}

@inproceedings{Hou,
author = {Hou, Kaiyuan and Xia, Stephen and Jiang, Xiaofan},
title = {BuMA: Non-Intrusive Breathing Detection using Microphone Array},
year = {2022},
isbn = {9781450394031},
publisher = {Association for Computing Machinery},
address = {New York, NY, USA},
url = {https://doi-org.proxy.lib.sfu.ca/10.1145/3539490.3539598},
doi = {10.1145/3539490.3539598},
booktitle = {Proceedings of the 1st ACM International Workshop on Intelligent Acoustic Systems and Applications},
pages = {1–6},
numpages = {6},
location = {Portland, OR, USA},
series = {IASA '22}
}

@article{Leonhardt,
author = {Leonhardt, Steffen and Leicht, Lennart and Teichmann, Daniel},
year = {2018},
month = {09},
pages = {3080},
title = {Unobtrusive Vital Sign Monitoring in Automotive Environments—A Review},
volume = {18},
journal = {Sensors},
doi = {10.3390/s18093080}
}

@unknown{Kim3,
author = {Kim, Dohun and Park, Hyukjin and Kim, Tonghyun and Kim, Wonjong},
year = {2023},
month = {08},
pages = {},
title = {Real-time Driver Monitoring System with Facial Landmark-based Eye Closure Detection and Head Pose Recognition},
doi = {10.21203/rs.3.rs-3223799/v1}
}

@inproceedings{Duri,
author = {Duri, Sastry and Gruteser, Marco and Liu, Xuan and Moskowitz, Paul and Perez, Ronald and Singh, Moninder and Tang, Jung-Mu},
title = {Framework for security and privacy in automotive telematics},
year = {2002},
isbn = {1581136005},
publisher = {Association for Computing Machinery},
address = {New York, NY, USA},
url = {https://doi-org.proxy.lib.sfu.ca/10.1145/570705.570711},
doi = {10.1145/570705.570711},
booktitle = {Proceedings of the 2nd International Workshop on Mobile Commerce},
pages = {25–32},
numpages = {8},
location = {Atlanta, Georgia, USA},
series = {WMC '02}
}

@article{Babaghayou,
author = {Babaghayou, Messaoud and Labraoui, Nabila and Adamou, Ado and Ari, A.A. and Lagraa, Nasreddine and Ferrag, Mohamed Amine},
year = {2020},
month = {10},
pages = {102618},
title = {Pseudonym change-based privacy-preserving schemes in vehicular ad-hoc networks: A survey},
volume = {55},
journal = {Journal of Information Security and Applications},
doi = {10.1016/j.jisa.2020.102618}
}

@inproceedings{Pape,
author = {Pape, Sebastian and Syed-Winkler, Sarah and Garcia, Armando Miguel and Chah, Badreddine and Bkakria, Anis and Hiller, Matthias and Walcher, Tobias and Lombard, Alexandre and Abbas-Turki, Abdeljalil and Yaich, Reda},
title = {A Systematic Approach for Automotive Privacy Management},
year = {2023},
isbn = {9798400704543},
publisher = {Association for Computing Machinery},
address = {New York, NY, USA},
url = {https://doi-org.proxy.lib.sfu.ca/10.1145/3631204.3631863},
doi = {10.1145/3631204.3631863},
booktitle = {Proceedings of the 7th ACM Computer Science in Cars Symposium},
articleno = {7},
numpages = {12},
location = {Darmstadt, Germany},
series = {CSCS '23}
}

@article{Dowthwaite,
title = {Privacy preferences in automotive data collection},
journal = {Transportation Research Interdisciplinary Perspectives},
volume = {24},
pages = {101022},
year = {2024},
issn = {2590-1982},
doi = {https://doi.org/10.1016/j.trip.2024.101022},
url = {https://www.sciencedirect.com/science/article/pii/S2590198224000083},
author = {Anna Dowthwaite and Dave Cook and Anna L. Cox}
}

@InProceedings{Walter,
author="Walter, Jonas
and Abendroth, Bettina",
editor="Zach{\"a}us, Carolin
and M{\"u}ller, Beate
and Meyer, Gereon",
title="Losing a Private Sphere? A Glance on the User Perspective on Privacy in Connected Cars",
booktitle="Advanced Microsystems for Automotive Applications 2017",
year="2018",
publisher="Springer International Publishing",
address="Cham",
pages="237--247",
isbn="978-3-319-66972-4"
}

@inproceedings{Nowara,
author = {Nowara, Ewa and Marks, Tim and Mansour, Hassan and Veeraraghavany, Ashok},
year = {2018},
month = {06},
pages = {1353-135309},
title = {SparsePPG: Towards Driver Monitoring Using Camera-Based Vital Signs Estimation in Near-Infrared},
doi = {10.1109/CVPRW.2018.00174}
}

@Article{Halin,
AUTHOR = {Halin, Anaïs and Verly, Jacques G. and Van Droogenbroeck, Marc},
TITLE = {Survey and Synthesis of State of the Art in Driver Monitoring},
JOURNAL = {Sensors},
VOLUME = {21},
YEAR = {2021},
NUMBER = {16},
ARTICLE-NUMBER = {5558},
URL = {https://www.mdpi.com/1424-8220/21/16/5558},
PubMedID = {34450999},
ISSN = {1424-8220},
DOI = {10.3390/s21165558}
}

@misc{Palao, title={Euro NCAP’s current and future in-cabin monitoring systems assessment}, url={https://research.chalmers.se/en/publication/536297}, journal={research.chalmers.se}, author={Palao, Adriano and Fredriksson, Rikard and Lenné, Mike}, year={2023}, month={Jun}}

@inproceedings{Tang,
author = {Tang, Mingyue and Teckchandani, Pranshu and He, Jizheng and Guo, Hanbo and Soltanaghai, Elahe},
title = {BSENSE: In-vehicle Child Detection and Vital Sign Monitoring with a Single mmWave Radar and Synthetic Reflectors},
year = {2024},
isbn = {9798400706974},
publisher = {Association for Computing Machinery},
address = {New York, NY, USA},
url = {https://doi-org.proxy.lib.sfu.ca/10.1145/3666025.3699352},
doi = {10.1145/3666025.3699352},
booktitle = {Proceedings of the 22nd ACM Conference on Embedded Networked Sensor Systems},
pages = {478–492},
numpages = {15},
location = {Hangzhou, China},
series = {SenSys '24}
}

@misc{DAD_Nguyen_2025, title={FTC takes action against General Motors for sharing drivers’ precise location and driving behavior data without consent}, url={https://www.ftc.gov/news-events/news/press-releases/2025/01/ftc-takes-action-against-general-motors-sharing-drivers-precise-location-driving-behavior-data}, journal={Federal Trade Commission}, author={DAD, Tom Koch (Acting and Nguyen, Stephanie T.}, year={2025}, month={Apr}}

@INPROCEEDINGS{Martin,
  author={Martin, Manuel and Roitberg, Alina and Haurilet, Monica and Horne, Matthias and Reiß, Simon and Voit, Michael and Stiefelhagen, Rainer},
  booktitle={2019 IEEE/CVF International Conference on Computer Vision (ICCV)}, 
  title={Drive\&Act: A Multi-Modal Dataset for Fine-Grained Driver Behavior Recognition in Autonomous Vehicles}, 
  year={2019},
  volume={},
  number={},
  pages={2801-2810},
  doi={10.1109/ICCV.2019.00289}}

@article{Hidalgo,
author = {Hidalgo-Muñoz, Antonio and Béquet, Adolphe and Astier-Juvenon, Mathis and Pepin, Guillaume and Fort, Alexandra and Jallais, Christophe and Tattegrain, Hélène and Gabaude, Catherine},
year = {2019},
month = {01},
pages = {525},
title = {Respiration and Heart Rate Modulation Due to Competing Cognitive Tasks While Driving},
volume = {12},
journal = {Frontiers in Human Neuroscience},
doi = {10.3389/fnhum.2018.00525}
}

@article{James,
title = {A systematic review of the use of in-vehicle telematics in monitoring driving behaviours},
journal = {Accident Analysis \& Prevention},
volume = {199},
pages = {107519},
year = {2024},
issn = {0001-4575},
doi = {https://doi.org/10.1016/j.aap.2024.107519},
url = {https://www.sciencedirect.com/science/article/pii/S0001457524000642},
author = {James Boylan and Denny Meyer and Won Sun Chen}
}

@article{Kafková,
author = {Kafková, Júlia and Babusiak, Branko and Pirnik, Rastislav and Kuchár, Pavol and Kekelak, Juraj and D’Ippolito, Filippo},
year = {2025},
month = {01},
pages = {115528},
title = {Seat to beat: Novel capacitive ECG integration for in-car cardiovascular measurement},
volume = {240},
journal = {Measurement},
doi = {10.1016/j.measurement.2024.115528}
}

@article{A_Malik,
  author = {A. Malik and J. Boger},
  title = {Zero-Effort Ambient Heart Rate Monitoring Using Ballistocardiography Detected Through a Seat Cushion: Prototype Development and Preliminary Study},
  year = {2021},
  journal = {JMIR Rehabilitation and Assistive Technologies},
  volume = {8},
  chapter = {e25996},
  url = {https://rehab.jmir.org/2021/2/e25996},
  doi = {10.2196/25996},
}

@misc{Bellan_2023, title={Harman’s driver-monitoring system can measure your heart rate}, url={https://techcrunch.com/2023/01/04/harmans-driver-monitoring-system-can-measure-your-heart-rate/}, journal={TechCrunch}, author={Bellan, Rebecca}, year={2023}, month={Jan}}

@article{Crandall,
author = {Crandall, David and Backstrom, Lars and Cosley, Dan and Suri, Siddharth and Huttenlocher, Daniel and Kleinberg, Jon},
year = {2010},
month = {12},
pages = {22436-41},
title = {Inferring social ties from geographic coincidences},
volume = {107},
journal = {Proceedings of the National Academy of Sciences of the United States of America},
doi = {10.1073/pnas.1006155107}
}

@article{András,
title = {Privacy pitfalls of releasing in-vehicle network data},
journal = {Vehicular Communications},
volume = {39},
pages = {100565},
year = {2023},
issn = {2214-2096},
doi = {https://doi.org/10.1016/j.vehcom.2022.100565},
url = {https://www.sciencedirect.com/science/article/pii/S2214209622001127},
author = {András Gazdag and Szilvia Lestyán and Mina Remeli and Gergely Ács and Tamás Holczer and Gergely Biczók}
}

@article{miller2015remote,
  title={Remote exploitation of an unaltered passenger vehicle},
  author={Miller, Charlie and Valasek, Chris},
  journal={Black Hat USA},
  volume={2015},
  number={S 91},
  pages={1--91},
  year={2015}
}

@article{Kim2,
title = {An effective automotive forensic technique utilizing various logs of Android-based In-vehicle infotainment systems},
journal = {Forensic Science International: Digital Investigation},
volume = {55},
pages = {301990},
year = {2025},
issn = {2666-2817},
doi = {https://doi.org/10.1016/j.fsidi.2025.301990},
url = {https://www.sciencedirect.com/science/article/pii/S2666281725001301},
author = {Sunjae Kim and Jeehun Jung and Haein Kang and Yejin Yoon and Seong-je Cho and Minkyu Park and Sangchul Han}
}

@inproceedings {Haohuang,
author = {Haohuang Wen and Qi Alfred Chen and Zhiqiang Lin},
title = {{Plug-N-Pwned}: Comprehensive Vulnerability Analysis of {OBD-II} Dongles as A New {Over-the-Air} Attack Surface in Automotive {IoT}},
booktitle = {29th USENIX Security Symposium (USENIX Security 20)},
year = {2020},
isbn = {978-1-939133-17-5},
pages = {949--965},
url = {https://www.usenix.org/conference/usenixsecurity20/presentation/wen},
publisher = {USENIX Association},
month = aug
}

@ARTICLE{Strandberg,
  author={Strandberg, Kim and Nowdehi, Nasser and Olovsson, Tomas},
  journal={IEEE Transactions on Intelligent Vehicles}, 
  title={A Systematic Literature Review on Automotive Digital Forensics: Challenges, Technical Solutions and Data Collection}, 
  year={2023},
  volume={8},
  number={2},
  pages={1350-1367},
  doi={10.1109/TIV.2022.3188340}}

@article{Sweeney,
author = {Sweeney, Latanya},
title = {k-anonymity: a model for protecting privacy},
year = {2002},
issue_date = {October 2002},
publisher = {World Scientific Publishing Co., Inc.},
address = {USA},
volume = {10},
number = {5},
issn = {0218-4885},
url = {https://doi-org.proxy.lib.sfu.ca/10.1142/S0218488502001648},
doi = {10.1142/S0218488502001648},
journal = {Int. J. Uncertain. Fuzziness Knowl.-Based Syst.},
month = oct,
pages = {557–570},
numpages = {14}
}

@INPROCEEDINGS{Machanavajjhala,
  author={Machanavajjhala, A. and Gehrke, J. and Kifer, D. and Venkitasubramaniam, M.},
  booktitle={22nd International Conference on Data Engineering (ICDE'06)}, 
  title={L-diversity: privacy beyond k-anonymity}, 
  year={2006},
  volume={},
  number={},
  pages={24-24},
  doi={10.1109/ICDE.2006.1}}

@article{Dwork,
author = {Dwork, Cynthia and McSherry, Frank and Nissim, Kobbi and Smith, Adam},
year = {2017},
month = {05},
pages = {17-51},
title = {Calibrating Noise to Sensitivity in Private Data Analysis},
volume = {7},
journal = {Journal of Privacy and Confidentiality},
doi = {10.29012/jpc.v7i3.405}
}

@article{Muhammad,
title = {Privacy preserving and data publication for vehicular trajectories with differential privacy},
journal = {Measurement},
volume = {173},
pages = {108675},
year = {2021},
issn = {0263-2241},
doi = {https://doi.org/10.1016/j.measurement.2020.108675},
url = {https://www.sciencedirect.com/science/article/pii/S0263224120311866},
author = {Muhammad Arif and Jianer Chen and Guojun Wang and Oana Geman and Valentina Emilia Balas}
}

@INPROCEEDINGS{Bhargavan,
  author={Bhargavan, Karthikeyan and Blanchet, Bruno and Kobeissi, Nadim},
  booktitle={2017 IEEE Symposium on Security and Privacy (SP)}, 
  title={Verified Models and Reference Implementations for the TLS 1.3 Standard Candidate}, 
  year={2017},
  volume={},
  number={},
  pages={483-502},
  doi={10.1109/SP.2017.26}}

@ARTICLE{Khodaei,
  author={Khodaei, Mohammad and Jin, Hongyu and Papadimitratos, Panagiotis},
  journal={IEEE Transactions on Intelligent Transportation Systems}, 
  title={SECMACE: Scalable and Robust Identity and Credential Management Infrastructure in Vehicular Communication Systems}, 
  year={2018},
  volume={19},
  number={5},
  pages={1430-1444},
  doi={10.1109/TITS.2017.2722688}}

@ARTICLE{Brecht,
  author={Brecht, Benedikt and Therriault, Dean and Weimerskirch, André and Whyte, William and Kumar, Virendra and Hehn, Thorsten and Goudy, Roy},
  journal={IEEE Transactions on Intelligent Transportation Systems}, 
  title={A Security Credential Management System for V2X Communications}, 
  year={2018},
  volume={19},
  number={12},
  pages={3850-3871},
  doi={10.1109/TITS.2018.2797529}}

@InProceedings{Bella,
author="Bella, Giampaolo
and Biondi, Pietro
and Tudisco, Giuseppe",
editor="Fischer-H{\"u}bner, Simone
and Lambrinoudakis, Costas
and Kotsis, Gabriele
and Tjoa, A. Min
and Khalil, Ismail",
title="Car Drivers' Privacy Concerns and Trust Perceptions",
booktitle="Trust, Privacy and Security in Digital Business",
year="2021",
publisher="Springer International Publishing",
address="Cham",
pages="143--154",
isbn="978-3-030-86586-3"
}

@inproceedings{Sleeper,
author = {Sleeper, Manya and Schnorf, Sebastian and Kemler, Brian and Consolvo, Sunny},
title = {Attitudes toward vehicle-based sensing and recording},
year = {2015},
isbn = {9781450335744},
publisher = {Association for Computing Machinery},
address = {New York, NY, USA},
url = {https://doi-org.proxy.lib.sfu.ca/10.1145/2750858.2806064},
doi = {10.1145/2750858.2806064},
booktitle = {Proceedings of the 2015 ACM International Joint Conference on Pervasive and Ubiquitous Computing},
pages = {1017–1028},
numpages = {12},
location = {Osaka, Japan},
series = {UbiComp '15}
}

@article{Sailesh,
title = {Measuring data sharing intention and its association with the acceptance of connected vehicles},
journal = {Transportation Research Part F: Traffic Psychology and Behaviour},
volume = {89},
pages = {423-436},
year = {2022},
issn = {1369-8478},
doi = {https://doi.org/10.1016/j.trf.2022.07.014},
url = {https://www.sciencedirect.com/science/article/pii/S1369847822001644},
author = {Sailesh Acharya and Michelle Mekker}
}

@article{Carlton,
author = {Carlton, Jason and Malik, Hafiz},
year = {2024},
month = {11},
pages = {198-212},
title = {A data privacy survey on personal identifiable information (PII) left on rental vehicle infotainment systems},
volume = {5},
journal = {Journal of Surveillance, Security and Safety},
doi = {10.20517/jsss.2024.07}
}

@inproceedings{Bossauer,
author = {Bossauer, Paul and Neifer, Thomas and Stevens, Gunnar and Pakusch, Christina},
title = {Trust versus Privacy: Using Connected Car Data in Peer-to-Peer Carsharing},
year = {2020},
isbn = {9781450367080},
publisher = {Association for Computing Machinery},
address = {New York, NY, USA},
url = {https://doi-org.proxy.lib.sfu.ca/10.1145/3313831.3376555},
doi = {10.1145/3313831.3376555},
booktitle = {Proceedings of the 2020 CHI Conference on Human Factors in Computing Systems},
pages = {1–13},
numpages = {13},
location = {Honolulu, HI, USA},
series = {CHI '20}
}

@article{Xusen,
title = {Investigating perceived risks and benefits of information privacy disclosure in IT-enabled ride-sharing},
journal = {Information \& Management},
volume = {58},
number = {6},
pages = {103450},
year = {2021},
note = {PACIS2019 Special Issue: Emerging technology, business, and applications in digital economy},
issn = {0378-7206},
doi = {https://doi.org/10.1016/j.im.2021.103450},
url = {https://www.sciencedirect.com/science/article/pii/S0378720621000240},
author = {Xusen Cheng and Tingting Hou and Jian Mou}
}

@article{Zhang,
author = {Zhang, Shikun and Feng, Yuanyuan and Bauer, Lujo and Cranor, Lorrie and Das, Anupam and Sadeh, Norman},
year = {2021},
month = {04},
pages = {282-304},
title = {“Did you know this camera tracks your mood?”: Understanding Privacy Expectations and Preferences in the Age of Video Analytics},
volume = {2021},
journal = {Proceedings on Privacy Enhancing Technologies},
doi = {10.2478/popets-2021-0028}
}

@inproceedings{Lee,
author = {Lee, Adam J. and Biehl, Jacob T. and Curry, Conor},
title = {Sensing or Watching? Balancing Utility and Privacy in Sensing Systems via Collection and Enforcement Mechanisms},
year = {2018},
isbn = {9781450356664},
publisher = {Association for Computing Machinery},
address = {New York, NY, USA},
url = {https://doi-org.proxy.lib.sfu.ca/10.1145/3205977.3205983},
doi = {10.1145/3205977.3205983},
booktitle = {Proceedings of the 23nd ACM on Symposium on Access Control Models and Technologies},
pages = {105–116},
numpages = {12},
location = {Indianapolis, Indiana, USA},
series = {SACMAT '18}
}

@article{Yao,
author = {Yao, Yaxing and Basdeo, Justin Reed and Mcdonough, Oriana Rosata and Wang, Yang},
title = {Privacy Perceptions and Designs of Bystanders in Smart Homes},
year = {2019},
issue_date = {November 2019},
publisher = {Association for Computing Machinery},
address = {New York, NY, USA},
volume = {3},
number = {CSCW},
url = {https://doi-org.proxy.lib.sfu.ca/10.1145/3359161},
doi = {10.1145/3359161},
journal = {Proc. ACM Hum.-Comput. Interact.},
month = nov,
articleno = {59},
numpages = {24}
}

@inproceedings{Kim,
author = {Kim, Injung and Lee, Adam J.},
title = {"I know what you did last semester": Understanding Privacy Expectations and Preferences in the Smart Campus},
year = {2024},
isbn = {9798400703300},
publisher = {Association for Computing Machinery},
address = {New York, NY, USA},
url = {https://doi-org.proxy.lib.sfu.ca/10.1145/3613904.3642174},
doi = {10.1145/3613904.3642174},
booktitle = {Proceedings of the 2024 CHI Conference on Human Factors in Computing Systems},
articleno = {980},
numpages = {15},
location = {Honolulu, HI, USA},
series = {CHI '24}
}

@inproceedings{Marky,
author = {Marky, Karola and Voit, Alexandra and St\"{o}ver, Alina and Kunze, Kai and Schr\"{o}der, Svenja and M\"{u}hlh\"{a}user, Max},
title = {”I don’t know how to protect myself”: Understanding Privacy Perceptions Resulting from the Presence of Bystanders in Smart Environments},
year = {2020},
isbn = {9781450375795},
publisher = {Association for Computing Machinery},
address = {New York, NY, USA},
url = {https://doi-org.proxy.lib.sfu.ca/10.1145/3419249.3420164},
doi = {10.1145/3419249.3420164},
booktitle = {Proceedings of the 11th Nordic Conference on Human-Computer Interaction: Shaping Experiences, Shaping Society},
articleno = {4},
numpages = {11},
location = {Tallinn, Estonia},
series = {NordiCHI '20}
}

@inproceedings{Emami,
author = {Emami-Naeini, Pardis and Bhagavatula, Sruti and Habib, Hana and Degeling, Martin and Bauer, Lujo and Cranor, Lorrie Faith and Sadeh, Norman},
title = {Privacy expectations and preferences in an IoT world},
year = {2017},
isbn = {9781931971393},
publisher = {USENIX Association},
address = {USA},
booktitle = {Proceedings of the Thirteenth USENIX Conference on Usable Privacy and Security},
pages = {399–412},
numpages = {14},
location = {Santa Clara, CA, USA},
series = {SOUPS '17}
}

@article{Das,
author = {Das, Anupam and Degeling, Martin and Smullen, Daniel and Sadeh, Norman},
year = {2018},
month = {07},
pages = {35-46},
title = {Personalized Privacy Assistants for the Internet of Things: Providing Users with Notice and Choice},
volume = {17},
journal = {IEEE Pervasive Computing},
doi = {10.1109/MPRV.2018.03367733}
}

@ARTICLE{Chow,
  author={Chow, Richard},
  journal={IEEE Security \& Privacy}, 
  title={The Last Mile for IoT Privacy}, 
  year={2017},
  volume={15},
  number={6},
  pages={73-76},
  doi={10.1109/MSP.2017.4251118}}

@inproceedings{Colnago,
author = {Colnago, Jessica and Feng, Yuanyuan and Palanivel, Tharangini and Pearman, Sarah and Ung, Megan and Acquisti, Alessandro and Cranor, Lorrie Faith and Sadeh, Norman},
title = {Informing the Design of a Personalized Privacy Assistant for the Internet of Things},
year = {2020},
isbn = {9781450367080},
publisher = {Association for Computing Machinery},
address = {New York, NY, USA},
url = {https://doi-org.proxy.lib.sfu.ca/10.1145/3313831.3376389},
doi = {10.1145/3313831.3376389},
booktitle = {Proceedings of the 2020 CHI Conference on Human Factors in Computing Systems},
pages = {1–13},
numpages = {13},
location = {Honolulu, HI, USA},
series = {CHI '20}
}

@inproceedings{Bermejo,
author = {Bermejo Fernandez, Carlos and Nurmi, Petteri and Hui, Pan},
title = {Seeing is Believing? Effects of Visualization on Smart Device Privacy Perceptions},
year = {2021},
isbn = {9781450386517},
publisher = {Association for Computing Machinery},
address = {New York, NY, USA},
url = {https://doi-org.proxy.lib.sfu.ca/10.1145/3474085.3475552},
doi = {10.1145/3474085.3475552},
booktitle = {Proceedings of the 29th ACM International Conference on Multimedia},
pages = {4183–4192},
numpages = {10},
location = {Virtual Event, China},
series = {MM '21}
}

@conference {Hosub,
	title = {Understanding User Privacy in Internet of Things Environments},
	booktitle = {2016 IEEE 3rd World Forum on Internet of Things (WF-IoT)},
	year = {2016},
	month = {Dec. 12-14},
	pages = {407-412},
	publisher = {IEEE Press},
	organization = {IEEE Press},
	address = {Reston, VA},
	doi = {10.1109/WF-IoT.2016.7845392},
	author = {Hosub Lee and Alfred Kobsa}
}

@book{Fleiss,
  title={Statistical Methods for Rates and Proportions},
  author={Fleiss, J.L. and Levin, B. and Paik, M.C.},
  isbn={9781118625613},
  series={Wiley Series in Probability and Statistics},
  url={https://books.google.ca/books?id=9VefO7a8GeAC},
  year={2013},
  publisher={Wiley}
}

@inproceedings{Choe,
author = {Choe, Eun Kyoung and Consolvo, Sunny and Jung, Jaeyeon and Harrison, Beverly and Kientz, Julie A.},
title = {Living in a glass house: a survey of private moments in the home},
year = {2011},
isbn = {9781450306300},
publisher = {Association for Computing Machinery},
address = {New York, NY, USA},
url = {https://doi-org.proxy.lib.sfu.ca/10.1145/2030112.2030118},
doi = {10.1145/2030112.2030118},
booktitle = {Proceedings of the 13th International Conference on Ubiquitous Computing},
pages = {41–44},
numpages = {4},
location = {Beijing, China},
series = {UbiComp '11}
}

@ARTICLE{Xun,
  author={Xun, Yijie and Liu, Jiajia and Kato, Nei and Fang, Yongqiang and Zhang, Yanning},
  journal={IEEE Transactions on Industrial Informatics}, 
  title={Automobile Driver Fingerprinting: A New Machine Learning Based Authentication Scheme}, 
  year={2020},
  volume={16},
  number={2},
  pages={1417-1426},
  doi={10.1109/TII.2019.2946626}}

@book{Westin1967PrivacyFreedom,
  title={Privacy and Freedom},
  author={Westin, A.F.},
  url={https://books.google.ca/books?id=EqGAfBTQreMC},
  publisher = {Atheneum},
  year={1968}
}

@misc{atockar_2014, title={Riding with the Stars: Passenger privacy in the NYC taxicab dataset}, url={https://agkn.wordpress.com/2014/09/15/riding-with-the-stars-passenger-privacy-in-the-nyc-taxicab-dataset/}, journal={Riding with the Stars: Passenger Privacy in the NYC Taxicab Dataset –}, author={atockar}, year={2014}, month={Sep}}

@article{Nissenbaum2004ContextualIntegrity,
  author  = {Helen Nissenbaum},
  title   = {Privacy as Contextual Integrity},
  journal = {Washington Law Review},
  volume  = {79},
  number  = {1},
  pages   = {119--158},
  year    = {2004}
}

@book{altman1975environment,
  title={The Environment and Social Behavior: Privacy, Personal Space, Territory, Crowding},
  author={Altman, I.},
  isbn={9780818501685},
  lccn={75014724},
  url={https://books.google.ca/books?id=GLBPAAAAMAAJ},
  year={1975},
  publisher={Brooks/Cole Publishing Company}
}

@inproceedings{Bloom,
author = {Bloom, Cara and Tan, Joshua and Ramjohn, Javed and Bauer, Lujo},
title = {Self-driving cars and data collection: privacy perceptions of networked autonomous vehicles},
year = {2017},
isbn = {9781931971393},
publisher = {USENIX Association},
address = {USA},
booktitle = {Proceedings of the Thirteenth USENIX Conference on Usable Privacy and Security},
pages = {357–375},
numpages = {19},
location = {Santa Clara, CA, USA},
series = {SOUPS '17}
}

@article{Jon_L1,
 ISSN = {03637425},
 URL = {http://www.jstor.org/stable/259124},
 author = {Jon L. Pierce and Tatiana Kostova and Kurt T. Dirks},
 journal = {The Academy of Management Review},
 number = {2},
 pages = {298--310},
 publisher = {Academy of Management},
 title = {Toward a Theory of Psychological Ownership in Organizations},
 urldate = {2025-09-11},
 volume = {26},
 year = {2001}
}

@article{Jon_L,
author = {Jon L. Pierce and Tatiana Kostova and Kurt T. Dirks},
title ={The State of Psychological Ownership: Integrating and Extending a Century of Research},

journal = {Review of General Psychology},
volume = {7},
number = {1},
pages = {84-107},
year = {2003},
doi = {10.1037/1089-2680.7.1.84},
URL = {    
        https://doi.org/10.1037/1089-2680.7.1.84
},
eprint = {    
        https://doi.org/10.1037/1089-2680.7.1.84}
}

@article{Cichy,
    author = {Cichy, Patrick and Salge, Torsten Oliver and Kohli, Rajiv},
    title = {Privacy Concerns and Data Sharing in the Internet of Things: Mixed Methods Evidence from Connected Cars},
    journal = {Management Information Systems Quarterly},
    volume = {45},
    number = {4},
    pages = {1863-1891},
    year = {2021},
    month = {12},
    issn = {0276-7783},
    doi = {10.25300/MISQ/2021/14165},
    url = {https://doi.org/10.25300/MISQ/2021/14165},
}

@article{Brandimarte,
author = {Brandimarte, Laura and Acquisti, Alessandro},
year = {2013},
month = {05},
pages = {340-347},
title = {Misplaced Confidences Privacy and the Control Paradox},
volume = {4},
journal = {Social Psychological and Personality Science},
doi = {10.1177/1948550612455931}
}
\clearpage
\appendix

\end{document}


\maketitle

This supplementary document provides additional methodological details referenced in the main paper, including (1) a summary of the interview topics and representative prompts used across both interview phases, and (2) a condensed summary of the final codebook structure.

\section{Summary of Interview Topics and Representative Prompts}
\label{supp:interview-guide}

\subsection{Phase 1: General Understanding and Calibration}
\label{supp:phase1}

At the beginning of each interview, participants were introduced to the concept of smart cabin sensing in order to establish a shared baseline understanding. We explained that a smart cabin uses in-cabin sensors, computing, and software to enhance safety, comfort, convenience, personalization, and other vehicle functions by sensing the cabin environment and occupants' states and behaviors.

Participants were then introduced to common categories of in-cabin sensors, including:
\begin{enumerate}[leftmargin=1.5em]
    \item \textbf{In-cabin cameras}, typically mounted around the rearview mirror, windshield header, A-pillar, steering column, or center display area, which may capture facial images, eye status, head pose, expressions, and gestures;
    \item \textbf{3D ToF cameras}, typically mounted in the dashboard, roof console, or overhead area, which may collect depth and posture information;
    \item \textbf{Millimeter-wave radar}, typically mounted in the roof liner, overhead console, or ceiling area, which may detect breathing, heart rate, and the presence of left-behind passengers or pets;
    \item \textbf{Ultrasonic sensors}, typically integrated near the dashboard, overhead area, or interior trim, which may detect passenger or object presence;
    \item \textbf{Seat pressure sensors}, typically embedded under the seat cushion or within the seat base, which may estimate occupant weight and presence;
    \item \textbf{Seatbelt sensors}, typically integrated into the seatbelt buckle or retractor assembly, which may detect fastening status;
    \item \textbf{Steering wheel capacitive and angle sensors}, typically embedded in the steering wheel rim, hub, or steering column assembly, which may detect hands on the wheel and steering movements;
    \item \textbf{In-cabin microphones}, typically mounted in the overhead console, headliner, near doors, or around the dashboard area, which may capture voice commands, conversations, and environmental sounds; and
    \item \textbf{Environmental sensors}, typically placed in the dashboard, HVAC module, center console, or roof area, which may monitor temperature, humidity, and air quality.
\end{enumerate}

For each type of sensor, participants were given a brief explanation of its typical location, the kinds of data it may collect, and its intended function. We then introduced the concept of \textit{privacy preferences} as users' different expectations and concerns regarding what kinds of data these sensors collect and how such data may be handled across different driving and riding situations.

\subsection{Phase 2: Scenario-Based Discussion}
\label{supp:phase2}

In the second phase, participants were asked to begin from one of three seeded scenarios. They were encouraged to reflect on each situation from multiple perspectives, including as a driver or passenger and as an owner or non-owner, and were also invited to introduce additional scenarios from their own experience when relevant. The three seeded scenarios were:

\begin{description}[leftmargin=1.5em, style=nextline]
    \item[\textbf{Personal Multi-Purpose Vehicle.}] You own a vehicle equipped with smart cabin sensors. You primarily use the car for family activities, such as driving your children to school or commuting to work, and occasionally for ride-hailing services. The smart cabin sensors are active throughout daily use.
    
    \item[\textbf{Rental Vehicle.}] You rent a vehicle equipped with smart cabin sensors. While you are unsure of the exact sensors, you know they are always active. You use this car for commuting during a trip.
    
    \item[\textbf{Ride-Hailing.}] You take a ride-hailing service for a commute. The vehicle's smart cabin sensors remain active throughout the trip.
\end{description}

For each scenario, we used the following core prompts iteratively, often referring to specific sensors such as the camera or microphone:
\begin{enumerate}[leftmargin=1.5em]
    \item In this specific scenario, what aspect of the data collected by this sensor concerns you the most (e.g., conversations, expressions, location, or driving state)?
    
    \item Can you describe specific situations in this scenario in which collecting this kind of data would make you feel uncomfortable, awkward, or worried? Why?
    
    \item In this scenario, would you want to adjust the functionality of this sensor (e.g., lower its sensitivity) or restrict it from collecting certain kinds of data? Under what conditions would you be most likely to do this?
    
    \item How do you balance the specific benefits provided by this sensor (e.g., fatigue monitoring, voice assistance, personalization) against the potential privacy risks in this scenario? When does the trade-off feel worthwhile, and when does it not?
\end{enumerate}

In addition to these core prompts, the interviewer used semi-structured follow-up questions to elicit concrete examples, clarify participants' reasoning, and probe how their preferences shifted across roles, relationships, and use contexts.

\section{Condensed Summary of the Final Codebook Structure}
\label{supp:codebook}

The full codebook contained 99 unique codes, which were iteratively developed and grouped into 9 higher-level themes through cross-identity comparative analysis. This supplementary material reports representative lower-level codes rather than the complete codebook, in order to show how the final analytic structure was organized without reproducing the entire coding inventory.

\setlength{\LTleft}{0pt}
\setlength{\LTright}{0pt}
\footnotesize
\renewcommand{\arraystretch}{1.15}

\begin{longtable}{L{2.1cm} L{3.0cm} L{4.0cm} L{1.6cm} L{3.9cm} L{0.8cm}}
\caption{Representative excerpt from the final codebook structure.}
\label{tab:supp-codebook} \\

\toprule
\textbf{Theme} & \textbf{Theme Description} & \textbf{Representative lower-level codes / subcodes} & \textbf{Primarily associated with} & \textbf{Example participants connected to the theme} & \textbf{Count} \\
\midrule
\endfirsthead

\toprule
\textbf{Theme} & \textbf{Theme Description} & \textbf{Representative lower-level codes / subcodes} & \textbf{Primarily associated with} & \textbf{Example participants connected to the theme} & \textbf{Count} \\
\midrule
\endhead

\bottomrule
\endfoot

\makecell[tl]{\textbf{C1.}\\\textbf{Presence of}\\\textbf{Companion}}
& How the presence of other occupants makes the cabin feel less fully private and changes users' privacy preferences and behavior.
& semi-private cabin; behavior moderation in transit; shielding personal smartphone screens; suppressing conversations in transit; protecting companions' privacy
& Owners \& Non-owners
& P1--P18
& 18 \\

\makecell[tl]{\textbf{C2.}\\\textbf{Perceived Trust}\\\textbf{in Travel}\\\textbf{Companions}}
& How familiarity with and trust in current co-passengers reduces or reshapes privacy concern.
& feeling secure with familiar co-passengers; trust in travel partners; greater comfort with trusted companions; accountability with less familiar co-passengers
& Owners \& Non-owners
& P1, P2, P3, P6, P8, P11, P15, P16, P17, P18
& 10 \\

\makecell[tl]{\textbf{C3.}\\\textbf{Social}\\\textbf{Pressure}}
& How companions' preferences, expectations, or possible reactions create interpersonal pressure around sensing and privacy decisions.
& differing privacy preferences among co-travelers; going along with companions' preferences; avoiding sensor-equipped vehicles because of companions; having to explain or justify sensors to passengers; feeling judged by sensor alerts
& Owners \& Non-owners
& P3, P4, P5, P6, P8, P9, P11, P13, P14, P15, P16, P17
& 12 \\

\makecell[tl]{\textbf{C4.}\\\textbf{Data Type}}
& How privacy concern varies depending on what kind of data is collected and what can be inferred from it.
& cameras and microphones as highly intrusive; biometric / physiological data as sensitive; air-quality / seat-pressure data as less sensitive; concern about third-party access to specific data types
& Owners \& Non-owners
& P1, P4, P5, P7, P8, P10, P11, P13, P15, P16
& 10 \\

\makecell[tl]{\textbf{C5.}\\\textbf{Cost-Benefit}\\\textbf{Trade-off}}
& How users weigh privacy concerns against perceived benefits such as safety, accountability, convenience, or financial incentives.
& privacy-for-safety trade-off; privacy-for-accountability / evidence; accepting driving-data collection for insurance or incident handling; benefit-dependent acceptance of sensing
& Owners \& Non-owners
& P10, P11, P13
& 3 \\

\makecell[tl]{\textbf{O1.}\\\textbf{Single-User or}\\\textbf{Shared Vehicle}}
& How owners' privacy preferences shift depending on whether the car is used only by them or regularly shared with others.
& shared-driver access concern; privacy in shared-use settings; concern about co-users accessing sensor data; private moments in shared cars
& Owners
& P2, P3, P10, P12, P14
& 5 \\

\makecell[tl]{\textbf{O2.}\\\textbf{Perceived Data}\\\textbf{Accuracy}}
& How owners' concern changes when they believe sensor data can be accurately linked to them or used to infer personal characteristics.
& combining multiple sensor streams to infer personal traits; concern about non-driving inferences; shared-driver noise as protection; sole-driver traceability; more detailed inference when only one person regularly drives the car
& Owners
& P2, P3, P6, P10, P12, P13, P14
& 7 \\

\makecell[tl]{\textbf{NO1.}\\\textbf{Perceived Trust}\\\textbf{in Absent}\\\textbf{Stakeholders}}
& How non-owners' privacy preferences depend on whether they trust stakeholders not currently in the car, such as owners, employers, platforms, or later users.
& trust in car owner; concern about employers / platforms; concern about future occupants or subsequent renters; limited perceived ownership of one's own data
& Non-owners
& P2, P4, P7, P8, P10, P11, P14, P15, P16
& 9 \\

\makecell[tl]{\textbf{NO2.}\\\textbf{Length of Car}\\\textbf{Possession}}
& How non-owners' privacy concern changes as their time with the vehicle increases and the space feels less temporary.
& short-trip manageability; long-trip self-censorship burden; accumulation of sensor traces over time; rental/work vehicle concern; public-to-private shift over time
& Non-owners
& P4, P7, P11, P14, P15, P16, P17
& 7 \\

\end{longtable}